# The Dirac-Coulomb Problem: a mathematical revisit

**A. D. Alhaidari [1], H. Bahlouli [1,2] and M. E. H. Ismail [3,4]**

[1] *Saudi Center for Theoretical Physics, Dhahran, Saudi Arabia*
[2] *Department of Physics, King Fahd University of Petroleum and Minerals, Dhahran 31261, Saudi Arabia*
[3] *Department of Mathematics, University of Central Florida, Orlando 32816, Florida*
[4] *King Saud University, Riyadh, Saudi Arabia*

**Abstract**: We obtain a symmetric tridiagonal matrix representation of the Dirac-Coulomb operator in a suitable complete square integrable basis. Orthogonal polynomials techniques along with Darboux method are used to obtain the bound states energy spectrum, the relativistic scattering amplitudes and phase shifts from the asymptotic behavior of the polynomial solutions associated with the resulting three-term recursion relation.

The Dirac equation was formulated by Paul Dirac in 1928 [1]. It is the most frequently used wave equation for the description of particle dynamics in relativistic quantum mechanics. It is a relativistically covariant first order linear matrix differential equation in space and time. It describes a spinor particle at relativistic energies below the threshold of pair production. It also embodies the features of quantum mechanics as well as special relativity. Moreover, the physics and mathematics of the Dirac equation are very rich, illuminating and provides a theoretical framework for different physical phenomena that are not present in the nonrelativistic regime. In fact, the Dirac equation predicted the existence of particle spin and required the existence of antiparticles. Moreover, it actually predated their experimental detection, making the discovery of the positron (the antiparticle of the electron) one of the greatest triumphs of modern theoretical physics. However, despite its fundamental importance and all the work that has been done over the years on this equation, its exact solution has been limited to a very small set of potentials.

Immediately after its discovery the Dirac equation was applied to the very fundamental problem that describes the interaction of an electron with a nucleus of charge $Z$ (in units of electron charge). This problem shed light on the importance of relativistic effects on the Hydrogen spectrum and was named the Dirac-Coulomb problem. The study of this problem resulted in a fine structure splitting of the energy levels due to spin-orbit coupling, a substantial relativistic contribution compared to the nonrelativistic description of the hydrogen atom [2]. For elements with high atomic number, $Z$, the relativistic effects become more pronounced, especially so for $s$ electrons that move at relativistic velocities as they penetrate the screening electrons near the core of high $Z$ atoms. In fact, even with nonrelativistic quantum mechanics, any atom with an atomic number greater than 137 would require its 1s electrons to be traveling faster than the speed of light. Using the Dirac equation, which accounts for the relativistic effects, it was shown that the wavefunction of the electron for atoms with $Z > 137$ is oscillatory and unbounded. The explanation of this mistery is a subject of interest to the physics community and one of the authors recently made a contribution in this regard [3].

It is the objective of our present work to use the lessons learned from the theory and techniques of orthogonal polynomials to obtain the bound states energy spectrum and relativistic scattering amplitudes and phase shifts from the resulting three-term recursion relation and its asymptotic solutions. Using the tridiagonal physics approach we will derive a three-term recursion relation for the expansion coefficients of the wave function. This second order recursion relation has two independent solutions that will be denoted by $P_n(x)$ and $P_n^*(x)$ with the help of which we can define the associated Green function as

$$G(z) = \lim_{n \to \infty} \frac{P_n^*(z)}{P_n(z)} = \int_{-\infty}^{+\infty} \frac{d\mu(t)}{z-t} \quad ; \quad \int_{-\infty}^{+\infty} P_n(x) P_m(x) d\mu(x) = \lambda_n \delta_{n,m}, \tag{1}$$

where $\mu(x)$ is the associated orthogonality measure. The orthogonal polynomials $P_n(x)$ and $P_n^*(x)$ are solutions of the same three-term recursion relation written in its symmetric form as

$$x P_n(x) = a_n P_n(x) + b_{n-1} P_{n-1}(x) + b_n P_{n+1}(x) \quad ; \quad n = 1,2,3\cdots, \tag{2}$$

but with different initial conditions [4]

$$P_0(x) = 1, \quad P_1(x) = \frac{x - a_0}{b_0} \quad ; P_0^*(x) = 0, \quad P_1^*(x) = \frac{1}{b_0}, \tag{3}$$

where the recursion coefficients $a_n$ and $b_n$ depend on $n$ and the potential parameters. The above choice of initial conditions ensures that the two solutions are independent. Thus, we devise an approach based on Darboux method to find the asymptotic behavior of each of the above polynomials for large values of $n$ that will then be used in the Green function (1). The associated measure can then be obtained from Stieljes inversion formula [4].

The Dirac equation for a free structureless spin $\frac{1}{2}$ particle of mass $m$ reads as follows:

$$\left( \gamma^\mu p_\mu - mc \right) \Psi = \left( i\hbar \gamma^\mu \partial_\mu - mc \right) \Psi = 0, \tag{4}$$

where $c$ is the speed of light and $\{\gamma^\mu\}_{\mu=0}^{3}$ are 4×4 constant matrices satisfying the anti-commutation relation $\{\gamma^\mu, \gamma^\nu\} = 2\mathcal{G}^{\mu\nu}$. The Mikowski space-time metric $\mathcal{G}$ = diag($+---$) and the 4-gradiant is $\partial_\mu = \left( \frac{1}{c} \frac{\partial}{\partial t}, \vec{\nabla} \right)$. Gauge invariant coupling of the spinor to the 4-vector electromagnetic potential $A_\mu$ is accomplished by the minimal substitution $p_\mu \to p_\mu - \frac{e}{c} A_\mu$, where $e$ is the electric charge of the Dirac particle. We write the vector potential $A^\mu = (V, c\vec{A})$ and the Dirac equation takes the following gauge invariant form

$$\left( i\hbar \gamma^\mu \partial_\mu - \frac{e}{c} \gamma^\mu A_\mu - mc \right) \Psi = 0, \tag{5}$$

For time independent potentials, we can write the total spinor wavefunction as $\Psi(t, \vec{r}) = e^{-i\mathcal{E}t/\hbar} \psi(\vec{r})$ giving

$$\left( \varepsilon + \frac{i\hbar}{mc} \vec{\alpha} \cdot \vec{\nabla} - \frac{e}{mc^2} V + \frac{e}{mc} \vec{\alpha} \cdot \vec{A} - \gamma_0 \right) \psi = 0, \tag{6}$$

where $\vec{\alpha} = \gamma^0 \vec{\gamma}$, $\varepsilon = \mathcal{E}/mc^2$. If we choose the standard matrix representation

$$\gamma^0 = \begin{pmatrix} I & 0 \\ 0 & -I \end{pmatrix}, \quad \vec{\gamma} = \begin{pmatrix} 0 & \vec{\sigma} \\ -\vec{\sigma} & 0 \end{pmatrix},$$

where $I$ is the 2×2 unit matrix and $\vec{\sigma}$ are the three 2×2 hermitian Pauli matrices. Then, we obtain for Eq. (6)

$$\begin{pmatrix} 1 + \frac{me}{\hbar^2}\lambdabar^2 V - \varepsilon & -\lambdabar\vec{\sigma}\cdot\left(i\vec{\nabla} + \frac{e}{\hbar}\vec{A}\right) \\ -\lambdabar\vec{\sigma}\cdot\left(i\vec{\nabla} + \frac{e}{\hbar}\vec{A}\right) & -1 + \frac{me}{\hbar^2}\lambdabar^2 V - \varepsilon \end{pmatrix} \begin{pmatrix} \psi^+ \\ \psi^- \end{pmatrix} = 0, \tag{6'}$$

with $\lambdabar = \hbar/mc$ is the Compton wavelength. For the Dirac-Coulomb problem $\vec{A} = 0$ and $V(r) = eZ/4\pi\epsilon_0 r$, where $Z$ is the dimension-less charge coupling. Thus, we can write $\frac{me}{\hbar^2}\lambdabar^2 V = (\lambdabar^2/a_0)\frac{Z}{r}$, where $a_0 = 4\pi\epsilon_0\hbar^2/me^2$ is the Bohr radius. Measuring length in units of $a_0$, the Dirac-Coulomb equation becomes

$$\begin{pmatrix} 1 + \lambdabar^2\frac{Z}{r} - \varepsilon & -i\lambdabar\vec{\sigma}\cdot\vec{\nabla} \\ -i\lambdabar\vec{\sigma}\cdot\vec{\nabla} & -1 + \lambdabar^2\frac{Z}{r} - \varepsilon \end{pmatrix} \begin{pmatrix} \psi^+(\vec{r}) \\ \psi^-(\vec{r}) \end{pmatrix} = 0. \tag{7}$$

Spherical symmetry enables one to separate the radial Dirac-Coulomb equation,

$$\begin{pmatrix} +1 + \lambdabar^2\frac{Z}{r} - \varepsilon & \lambdabar\left(\frac{\kappa}{r} - \frac{d}{dr}\right) \\ \lambdabar\left(\frac{\kappa}{r} + \frac{d}{dr}\right) & -1 + \lambdabar^2\frac{Z}{r} - \varepsilon \end{pmatrix} \begin{pmatrix} \chi^+(r) \\ \chi^-(r) \end{pmatrix} = 0, \tag{8}$$

where the spin-orbit quantum number $\kappa = \pm 1, \pm 2,..$ and it is related to the orbital angular momentum quantum number $\ell$ by $\kappa = \pm(\ell+1)$ for $\ell = j \pm \frac{1}{2}$. Equation (8) results in two coupled first order differential equations for the two radial spinor components $\chi^\pm$. To uncouple these equations such that the resulting second order differential equation for one of the two components become Schrödinger-like, we apply the following unitary transformation $\mathcal{U}(\xi) = \exp(\frac{i}{2}\xi\sigma_2)$ on Eq. (8), where $\xi$ is a real angular parameter and $\sigma_2 = \begin{pmatrix} 0 & -i \\ i & 0 \end{pmatrix}$. This requirement results in a constraint on the angular parameter $\xi$, which should satisfy: $\sin\xi = \pm\lambdabar Z/\kappa$. The top/bottom sign corresponds to the positive/negative energy solution space of the problem. Following these steps, the transformed Dirac equation will result in Schrödinger-like equation for one of the two components. The other component is obtained by substituting the solution of this equation into the "kinetic balance" relation, which is a first order differential relation obtained from the transformed Dirac equation and relating the two spinor components. For positive energy, this scheme transforms Eq. (8) into

$$\begin{pmatrix} \frac{\gamma}{\kappa} + 2\lambdabar^2\frac{Z}{r} - \varepsilon & \lambdabar\left(-\frac{Z}{\kappa} + \frac{\gamma}{r} - \frac{d}{dr}\right) \\ \lambdabar\left(-\frac{Z}{\kappa} + \frac{\gamma}{r} + \frac{d}{dr}\right) & -\frac{\gamma}{\kappa} - \varepsilon \end{pmatrix} \begin{pmatrix} \phi^+(r) \\ \phi^-(r) \end{pmatrix} = 0,$$

where the new spinor component are related to the old one by the unitary transformation, $\begin{pmatrix} \phi^+ \\ \phi^- \end{pmatrix} = e^{\frac{i}{2}\xi\sigma_2}\begin{pmatrix} \chi^+ \\ \chi^- \end{pmatrix}$ and $\gamma = \kappa\sqrt{1 - (\lambdabar Z/\kappa)^2}$. Eliminating the lower spinor component in favor of the upper gives the following Schrödinger-like equation

$$\left[-\frac{d^2}{dr^2} + \frac{\gamma(\gamma+1)}{r^2} + 2\frac{Z\varepsilon}{r} + \frac{\varepsilon^2 - 1}{\lambdabar^2}\right]\phi^+ = 0, \tag{9}$$

On the other hand, the lower component is given by the kinetic balance relation as
$$\phi^- = \frac{\lambda}{\varepsilon + \gamma/\kappa}\left(-\frac{Z}{\kappa} + \frac{\gamma}{r} + \frac{d}{dr}\right)\phi^+ . \tag{9'}$$

Now we select a basis set that enables us to obtain a tridiagonal representation for the Dirac-Coulomb Hamiltonian. One such useful basis, which is sometimes referred to as the "Laguerre basis", has elements for the upper spinor component of the form
$$\zeta_n^+(r) = A_n(\omega r)^{\gamma+1} e^{-\omega r/2} L_n^{2\gamma+1}(\omega r), \tag{10}$$
where $A_n$ is a normalization constant, $\omega$ is a positive inverse length scale parameter, and $L_n^\nu(x)$ is the associated Laguerre polynomial of order $n$. The lower spinor basis element, $\zeta_n^-(r)$, is related to the upper by the kinetic balance relation (9') [5]. This basis is chosen such that it produces a tridiagonal matrix representation for the Dirac-Coulomb Hamiltonian (Kinetic + Coulomb). Consequently, an exact regular solution of the problem is obtained in terms of orthogonal polynomials. These are solutions of the three-term recursion relation resulting from the tridiagonal structure of the matrix representation of the wave equation. The orthogonal polynomials appear in the expansion coefficients of the wavefunction, $\phi(r,\varepsilon) = \sum_{n=0}^{\infty} f_n(\varepsilon)\zeta_n(r,\varepsilon)$. After some manipulations, one obtains the following three-term recursion relation for the positive energy solution of the Dirac-Coulomb problem [5]
$$\left[a_n\left(\frac{\varepsilon^2-1-\beta^2}{\varepsilon^2-1+\beta^2}\right) - \frac{\alpha\varepsilon}{\varepsilon^2-1+\beta^2}\right]f_n(\varepsilon) = b_{n-1}f_{n-1}(\varepsilon) + b_n f_{n+1}(\varepsilon), \tag{11}$$
where $a_n = (n+\gamma+1)$, $b_n = \frac{1}{2}\sqrt{(n+1)(n+2\gamma+2)}$, $\alpha = \lambda^2\omega Z$, and $\beta = \frac{1}{2}\lambda\omega$. In this three-term recursion relation the relativistic energy is bounded ($|\varepsilon| \leq 1$) and discrete for bound states, while it is greater than unity, $|\varepsilon| \geq 1$ for scattering states where the spectrum is continuous. The above equation holds for $\kappa > 0$. A similar equation could be obtained for $\kappa < 0$ with the only difference is in the $n$-dependent recursion coefficients ($a_n$ and $b_n$) where the replacement $\gamma \to -\gamma-1$ is required. Our goal now is to find the two independent solutions associated with the recursion relation (11), then use Darboux method to find the asymptotic behavior associated with these polynomials which will then lead us to the explicit calculations of both bound states, scattering amplitude and phase shift.

To solve the above recursion relation we first recall that the Pollaczek polynomials, $P_n^\lambda(x;a,b)$, satisfy the following recursion relation [6]
$$[(n+\lambda+a)x + b]P_n^\lambda(x;a,b) = \tfrac{1}{2}(n+2\lambda-1)P_{n-1}^\lambda(x;a,b) + \tfrac{1}{2}(n+1)P_{n+1}^\lambda(x;a,b), \tag{12a}$$
with the initial relations
$$P_0^\lambda = 1,\ P_1^\lambda = 2(\lambda+a)x + 2b. \tag{12b}$$
To make a transparent comparison with our recursion relation (11), we bring the above recursion relation (12a) to a symmetric form through the following transformation
$$P_n^\lambda(x) = \sqrt{\frac{\Gamma(n+2\lambda)}{\Gamma(n+1)\Gamma(2\lambda+1)}} Q_n^\lambda(x),$$
which brings Eq. (12a) to

$$2[(n + \lambda + a)x + b]Q_n^\lambda(x;a,b) = \sqrt{n(n+2\lambda-1)}Q_{n-1}^\lambda(x;a,b) \tag{13}$$
$$+ \sqrt{(n+1)(n+2\lambda)}Q_{n+1}^\lambda(x;a,b)$$

Comparing this equation with our main recursion relation (11), we can make the following identifications

$$x = \frac{\varepsilon^2 - 1 - \beta^2}{\varepsilon^2 - 1 + \beta^2}, \quad a = 0, \quad b = -\frac{\alpha\varepsilon}{\varepsilon^2 - 1 + \beta^2}, \quad \lambda = \gamma + 1. \tag{14}$$

Thus the expansion coefficients, $f_n(\varepsilon)$, of the spinor wavefunction in the basis, $\{\zeta_n\}$, are related to the Pollaczek polynomials, $P_n^\lambda(x;a,b)$, for the above specific values of the parameters. The recurrence realtion (12a) and the initial condition (12b) show that $P_n^\lambda(x;a,b)$ is a polynomial of degree $n$, in $x$, in $a$, and in $b$. It is also clear that $(\varepsilon^2 - 1 + \beta^2)^n P_n^\lambda(x;a,b)$ is a polynomial of degree $2n$ in the energy variable $\varepsilon$, with $n = 0,1,2,\ldots$. Our task now is to obtain the asymptotic behavior of these polynomials using Darboux's method. For this purpose we start from the Pollaczek polynomials generating function (see section 4.8 in [6])

$$\sum_{n=0}^\infty P_n^\lambda(x;a,b)t^n = (1 - te^{i\theta})^{-\lambda + i\Phi(\theta)}(1 - te^{-i\theta})^{-\lambda - i\Phi(\theta)}, \tag{15}$$

where $x = \cos\theta$ ($\theta$ could be complex) and $\Phi(\theta) = (a\cos\theta + b)/\sin\theta = \frac{b}{\sin\theta}$. The asymptotic behavior is dominated by the smallest pole (poles) of the generating function on the right-hand side of (15) which, through the use of Darboux method (see theorem 1.2.4 in [6]), gives us the sought after asymptotic behavior ($n \to \infty$) of the polynomials. The following procedure will enable us to find this asymptotic behavior:
1. Locate the singularity of smallest modulus since it dominates the asymptotic behavior.
2. Compute the expansion of the generating function in the neighbourhood of this singularity.
3. Use this expansion to lead to the corresponding asymptotic behavior of the generating function expansion coefficients.

Application of the above algorithm leads to the subsequent results.

**Scattering states:** These states are represented by the energy continuum which, in our case, holds for $|\varepsilon| > 1$ and leads to $-1 < x < +1$. The generating function (15) has singularities at $e^{\pm i\theta}$, which can be defined in terms of the relativistic energy and our model parameters as follows

$$e^{\pm i\theta} = \frac{\sqrt{\varepsilon^2 - 1} \pm i\beta}{\sqrt{\varepsilon^2 - 1} \mp i\beta} = \frac{\sqrt{1-\varepsilon^2} \pm \beta}{\sqrt{1-\varepsilon^2} \mp \beta}. \tag{16}$$

Hence, $\theta$ is pure imaginary if $\varepsilon \in [-1,+1]$ and real if $\varepsilon \notin [-1,+1]$. Thus, for scattering states ($|\varepsilon| \geq 1$), $\theta$ is real and both poles of the generating function contribute to the asymptotic behavior since they both have unit magnitude. To obtain the asymptotic behavior of the Pollaczek polynomials we study the limit as $n \to \infty$ and identify the asymptotic behavior of (15) with that of the associated comparison function to give

$$P_n^\lambda(x;a,b) \approx \left[(1-e^{2i\theta})^{-\lambda+i\Phi}(\lambda+i\Phi)_n \frac{e^{-in\theta}}{n!} + cc\right], \tag{17}$$

where *cc* stands for complex conjugate and $(C)_n = C(C+1)(C+2)...(C+n-1) = \Gamma(n+C)/\Gamma(C)$. Since $\theta$ is real for scattering states both poles of the generating function at $e^{\pm i\theta}$ give contribution to (15), hence the existence of complex conjugation in (17). Using the asymptotic behavior of the Gamma function [6], $\frac{\Gamma(n+A)}{\Gamma(n+B)} n^{B-A} \approx 1$, we can rewrite (17) as follows

$$P_n^\lambda(x;a,b) \approx \left[ (1-e^{2i\theta})^{-\lambda+i\Phi} e^{-in\theta} \frac{n^{\lambda+i\Phi-1}}{\Gamma(\lambda+i\Phi)} + cc \right]. \tag{18}$$

At this stage we define $\Gamma(\lambda+i\Phi) = |\Gamma(\lambda+i\Phi)| e^{i\psi}$ where $\psi$ is a real number. Thus, the phase in the expression

$$\frac{n^{+i\Phi} e^{-in\theta}}{\Gamma(\lambda+i\Phi)} (1-e^{2i\theta})^{-\lambda+i\Phi} = \frac{n^{+i\Phi} e^{-in\theta} e^{-i\psi}}{|\Gamma(\lambda+i\Phi)|} e^{-i\theta(\lambda-i\Phi)} (2e^{-i\pi/2} \sin\theta)^{-(\lambda-i\Phi)},$$

can be written as $n\theta + \psi_n$, where $\psi_n = \psi + \lambda\left(\theta - \tfrac{1}{2}\pi\right) - \Phi(\theta)\ln(2n)$. On the other hand, the modulus of the leading term appearing in (17) can be written as follows

$$\left| (1-e^{2i\theta})^{-\lambda+i\Phi} e^{-in\theta} \frac{n^{\lambda+i\Phi-1}}{\Gamma(\lambda+i\Phi)} \right| = \frac{n^{\lambda-1}(2\sin\theta)^{-\lambda}}{|\Gamma(\lambda+i\Phi)|} e^{\left(\frac{\pi}{2}-\theta\right)\Phi}.$$

Combining these results give the desired asymptotic behavior

$$\left\{ P_n^\lambda(x;a,b) |\Gamma(\lambda+i\Phi)| (2\sin\theta)^\lambda e^{\left(\theta-\frac{\pi}{2}\right)\Phi} \right\} \approx 2n^{\lambda-1} \cos(n\theta + \psi_n)$$

On the other hand, the orthonormal Pollaczek polynomials are defined by

$$p_n^\lambda(x;a,b) = \sqrt{\frac{\Gamma(n+1)(\lambda+a+n)}{\Gamma(n+2\lambda)}} P_n^\lambda(x;a,b).$$

So that its asymptotic behavior becomes

$$p_n^\lambda(x;a,b) \approx n^{1-\lambda} P_n^\lambda(x;a,b) \approx \frac{2e^{\left(\frac{\pi}{2}-\theta\right)\Phi}}{|\Gamma(\lambda+i\Phi)|(2\sin\theta)^\lambda} \cos(n\theta + \psi_n). \tag{19}$$

In this form one can easily identify the phase shift, $\psi_n = \psi + \lambda\left(\theta - \tfrac{1}{2}\pi\right) - \Phi(\theta)\ln(2n)$, and the scattering amplitude as the energy dependent factor multiplying the geometric function. Usually, the argument of the geometric function in the asymptotic form of typical polynomials is linear in *n*. Here, however, the logarithmic *n*-dependence of the phase shift is reminiscent of a similar behavior in the configuration space wavefunction due to the long-range behavior of the Coulomb potential.

**Bound states:** In this situation we have $|\varepsilon|<1$ which results in $|x|>1$. Because the bound states are associated with discrete eigenvalues we expect that $f_n(\varepsilon)$ to be in $L^2$, whereas $f_n(\varepsilon) \notin L^2$ if $\varepsilon$ is not a bound state. So the main term in the large *n* asymptotic expansion of $f_n(\varepsilon)$ should vanish for bound states, this point is of fundamental importance to our approach. The parameters are still defined as

$$x = \cos\theta = \frac{\varepsilon^2 - 1 - \beta^2}{\varepsilon^2 - 1 + \beta^2}, \quad \sin\theta = \pm\sqrt{1-\cos^2\theta} = \pm\frac{2\beta\sqrt{\varepsilon^2-1}}{\varepsilon^2 - 1 + \beta^2}.$$

However, $\theta$ is pure imaginary in this case. To find the leading term in the asymptotic expansion (15) we need to treat each contribution separately. First we assume that $\left|e^{-i\theta}\right|<1<\left|e^{+i\theta}\right|$, with $e^{\pm i\theta}=x\pm\sqrt{x^2-1}$. We take $x>1$ and assume that $\sqrt{x^2}\approx x$ as $x\to\pm\infty$. Under these circumstance the leading term is given by

$$P_n^\lambda(x;a,b)\approx\left[(1-e^{-2i\theta})^{-\lambda-i\Phi}(\lambda-i\Phi)_n\frac{e^{in\theta}}{n!}\right]\approx n^{\lambda-i\Phi-1}e^{in\theta}\frac{(1-e^{-2i\theta})^{-\lambda-i\Phi}}{\Gamma(\lambda-i\Phi)}. \qquad (20)$$

The bound state requirement is that $\sum_{n=0}^{\infty}\left|u_n(x)\right|^2$ converges, where $\{u_n(x)\}$ are the orthogonal polynomials [7]. This means that on the bound states the asymptotic values of the polynomials are smaller than their asymptotic values in the neighboring points. Clearly in order for this to happen, the main term in the large $n$ asymptotics of $P_n^\lambda(x;a,b)$ must vanish. This requirement will then be met if $\frac{1}{\Gamma(\lambda-i\Phi)}=0$, as can be seen in (20), that is, $\lambda-i\Phi=-n$, where $n=0,1,2,\ldots$ Thus, solving the equation $-\Phi^2=(n+\lambda)^2$ in the energy variable leads to

$$\varepsilon_n=+\left[1+\left(\tfrac{\lambda Z}{n+\gamma+1}\right)^2\right]^{-1/2}, \qquad (21)$$

which agrees with the standard result obtained by direct solution of the Dirac equation [5] for the positive energy solution and $\kappa>0$. Note that we took the positive sign of the square root in (21) since we are dealing with the case of positive energy solutions. Now, let us calculate explicitly $e^{\pm i\theta}$ in terms of the bound states spectrum obtained in (21) and make sure that the assumed condition, $\left|e^{+i\theta}\right|>1>\left|e^{-i\theta}\right|$, holds true. For this purpose, we can rewrite the exponential function as follows

$$e^{i\theta}=\cos\theta+i\sin\theta=\frac{\left(\sqrt{\varepsilon_n^2-1}\pm i\beta\right)^2}{\varepsilon_n^2-1+\beta^2}=\frac{\sqrt{1-\varepsilon_n^2}+\beta}{\sqrt{1-\varepsilon_n^2}-\beta}, \qquad (22)$$

where the upper sign has been chosen in (16) to be compatible with the requirement $\left|e^{+i\theta}\right|>1>\left|e^{-i\theta}\right|$, putting this result back in the original condition gives

$$\Phi=-i(n+\lambda)=\frac{b}{\sin\theta}=i\frac{\alpha\varepsilon_n}{2\beta\sqrt{1-\varepsilon_n^2}},$$

which requires that $\alpha<0$ and thus $Z<0$ (i.e., attractive Coulomb potential) as a requirement for the existence of bound states. A similar analysis for $x<-1$ results in $\left|e^{+i\theta}\right|<1<\left|e^{-i\theta}\right|$ and give the asymptotic contribution

$$P_n^\lambda(x;a,b)\approx\left[(1-e^{+2i\theta})^{-\lambda+i\Phi}(\lambda+i\Phi)_n\frac{e^{-in\theta}}{n!}\right]\approx n^{\lambda+i\Phi-1}e^{-in\theta}\frac{(1-e^{2i\theta})^{-\lambda+i\Phi}}{\Gamma(\lambda+i\Phi)}, \qquad (23)$$

with the bound state condition $\lambda+i\Phi=-n$, where $n=0,1,2,\ldots$ resulting in the same energy spectrum formula for the bound state (21). However, the choice of signs and in the numerator and denominators of (22) must be exchanged to meet the requirement that $\left|e^{+i\theta}\right|<1<\left|e^{-i\theta}\right|$.

Using the general formula of the Pollaczek polynomials, we obtain the positive energy solution of the recursion for $\kappa>0$ as the expansion coefficients

$$f_n^\gamma(\varepsilon;\alpha,\beta)=\sqrt{\frac{\Gamma(2\gamma+2)}{\gamma+1}\frac{n+\gamma+1}{\Gamma(n+2\gamma+2)\Gamma(n+1)}}\,e^{in\theta}(\gamma+1-i\Phi(\theta))_n\,{}_2F_1\!\left(\begin{smallmatrix}-n,\,\gamma+1+i\Phi(\theta)\\-n-\gamma-1+i\Phi(\theta)\end{smallmatrix}\Big|e^{-2i\theta}\right).$$

For $\kappa < 0$, we just make the replacement $\gamma \to -\gamma - 1$. On the other hand, the negative energy solution is obtained from the positive energy solution by the map:

$$Z \to -Z, \quad \kappa \to -\kappa, \quad \varepsilon \to -\varepsilon, \text{ and } \zeta_n^+ \leftrightarrow \zeta_n^-. \tag{24}$$

In conclusion we would like to stress that our methodology is inspired by lessons learned from the theory of orthogonal polynomials and the associated Nevai theorem [8], which enables one to find the continuous piece of the energy spectrum from the asymptotic behavior. However, the Pollaczek polynomials do not fall in this category but it is surprising that the spirit of Nevai theorem still holds. Using orthogonal polynomials techniques along with the Darboux method, which enabled us to find the asymptotic behavior of the Pollaczek polynomials, we were able to obtain the bound states energy spectrum, the relativistic scattering amplitudes and phase shifts from the asymptotic behavior of the polynomial solutions associated with the resulting three-term recursion relation. Our results agree with those obtained by direct methods in the literature [5].

**Acknowledgements:** This work is sponsored by the Saudi Center for Theoretical Physics (SCTP). Partial support by King Fahd University of Petroleum and Minerals under group project number RG1108-1 and RG1108-2 is acknowledged. M. E. H. Ismail acknowledges NPST Program of King Saud University, Saudi Arabia, 10-MAT1293-02 for their support.